**Origin of Chirality in the Molecules of Life**

J. A. Cowan and R. J. Furnstahl

Contribution from the Department of Chemistry and Biochemistry, The Ohio State University, 100 West 18th Avenue, Columbus, Ohio 43210, and the Department of Physics, The Ohio State University, 191 West Woodruff Avenue, Columbus, Ohio 43210.

**Abstract**

Molecular chirality is inherent to biology and cellular chemistry. In this report, the origin of enantiomeric selectivity is analyzed from the viewpoint of the "RNA World" model, based on the autocatalytic self-replication of glyceraldehyde as a precursor for simple sugars, and in particular ribose, as promoted by the formose reaction. Autocatalytic coupling of formaldehyde and glycolaldehyde produces glyceraldehyde, which contains a chiral carbon center that is carried through in formation of the ribose ring. The parity non-conserving weak interaction is the only inherently handed property in nature and is herein shown to be sufficient to differentiate between two enantiomeric forms in an autocatalytic reaction performed over geologically relevant time scales, but only in the presence of a catalytic metal ion such as divalent calcium or higher Z alkaline earth elements. This work details calculations of the magnitude of the effect, the impact of various geologically-available metal ions, and the influence on evolution and dominance of chirality in the molecules of life.

## Introduction

The molecular understanding of life processes is based on the chemistry of molecules that frequently contain one or more chiral centers.[1] That is, molecular centers that possess the quality of mirror image symmetry. In most cases this symmetry stems from the relative positioning of substituents around a tetrahedral carbon center (Fig. 1), although chirality can also emerge from higher structural ordering as reflected in a helix or propeller shape.[2]

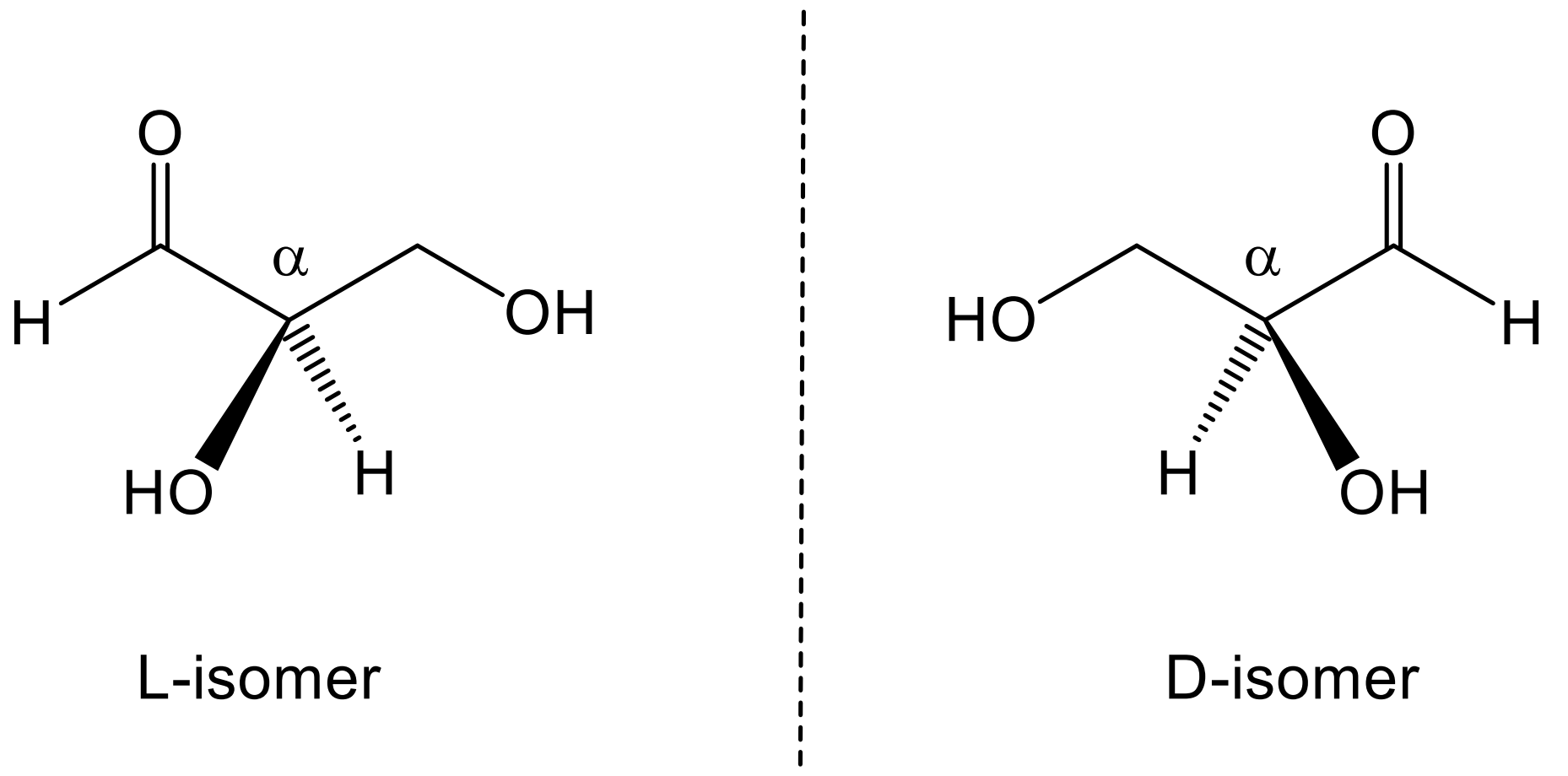

**Fig. 1.** Elements of molecular chirality illustrated through the two mirror-image enantiomers of glyceraldehyde.

Nature demonstrates a preference for left-handed symmetry, at least at the most fundamental level of molecular symmetry reflected by the molecular building blocks of life (nucleic acids and amino acids).[1] Once an inherent preference for one chiral form over another has been established, it is not difficult to understand, from the chemical principles of chiral induction and diastereomeric selection, how that chirality is propagated to downstream molecular products resulting from complex biosynthetic pathways,[3-6] often involving natural protein- or RNA-based catalysts with their own enantiomeric preferences. Consequently, the more fundamental question can be stated as, "how did an intrinsic preference for one chiral form initially arise?".

Prior work on this problem has been well detailed by Guijarro and Yus,[1] which provides a valuable overview of the primary mechanisms by which chirality in life molecules has been proposed. Both spontaneous symmetry breaking during crystallization,[7] and seeding mechanisms from extraterrestrial meteorites and comets[8,9] are possible pathways for initiation of chiral selectivity, but in all of these cases prior establishment of a specific enantiomeric form is required. Over the past decade, the major research emphasis has focused on the mechanisms and pathways that could promote amplification or selection of selected enantiomers by kinetic criteria.[10-16] The molecules under investigation include both natural biomolecules,[14,17] especially amino acids, as well as regular organic compounds.[4,6,11,12,15,18-21]

Experimentally, there are a number of principle kinetic factors that sustain autocatalytic generation of a selected enantiomer.[22-25] Based on an analysis of the Soai autocatalytic reaction, and the stochastic computational model of Kondepudi and Nelson, Hawbaker and Blackmond have calculated the energy requirement to break symmetry with a regular chiral bias to lie between 1.5 x $10^{-7}$ and 1.5 x $10^{-8}$ kJ/mole (5.7 x $10^{-11}$ and 5.7 x $10^{-12}$ au).[10] This is significantly higher than could be achieved through a parity non-conservation (PNC) type of mechanism, and whether the latter effect could promote enantiomeric selectivity on kinetic grounds has been questioned.[26] However, alternative pathways to achieve chiral selectivity have been identified. Kondepudi and Nelson have detailed the circumstances under which autocatalysis would provide a viable kinetic pathway to achieve such an outcome.[25,27,28] In terms of a reversible Frank-type model, autocatalysis provides a mechanism that allows one enantiomer to dominate even under conditions of random fluctuations. They estimated time frames of the order of up to 100,000 years for such selectivity to arise, but would depend on reaction rates, local concentrations, a continuous provision of starting materials, diffusion rates, reaction reversibility and diffusion rates. Such a mechanism does not, of course, have a specific enantiomeric bias, but in combination with an autocatalytic process to “seed” a preferred enantiomer, then chiral selectivity could be promoted via a PNC-type model. Furthermore, a simple stochastic model does not consider an ongoing “chiral prompt” that continues to be in effect throughout the period under consideration. This scenario has not previously been considered.

In this paper we demonstrate that given reasonable allowance for these factors, a PNC mechanism should meet the energy limitations for enantiomeric selectivity through autocatalysis, where PNC represents the primary source of chiral induction, if the autocatalytic reaction involves alkaline earth metal ions over a more extended period of time. Any assumptions made in these calculations are extremely conservative, to provide a lower limit on the expected magnitude of the effect.

Although the idea of the weak nuclear force allowing chiral discrimination has previously been put forward,[1,26,28-31] the molecules considered were relatively simple, and not clearly applicable to understanding the evolution of chirality in nature. Moreover, the effect was thought to be too small to observe in the absence of a heavy atom[32-35] but conditions to enhance the influence have been better established over time. These include application to heavy elements where the $Z^5$ dependence of the PNC effect is beneficial, as well as the involvement of more than one atom to avoid the single-center problem.[36,37] In early work, Faglioni and Lazerretti estimated $a = \Delta E^{PNC} \sim 4.9 \times 10^{-18}$ au for a single chiral center CHFClBr ,[38] while in a relativistic calculation of $H_2X_2$ (X = O, S, Se, Te and Po) Laerdahl and Scwerdtfeger estimated $\Delta E^{PNC}$ values ranging from $7.06 \times 10^{-19}$ au to $1.55 \times 10^{-12}$ au from the lighter O to heavier Po molecules.[39] In general, however, it is clear that any influence on the chirality of a carbon center that is dependent on the intrinsic influence of that carbon, or directly bonded low MW elements, will result in a $\Delta E^{PNC}$ that will be negligible in impacting enantiomeric selectivity.

Herein, we report the results and conclusions from an analysis of how preferential formation of a specific enantiomer of glyceraldehyde (supplementary file) could have resulted in broader establishment of chirality throughout the natural world. The model is based on an RNA-centered evolutionary pathway,[40,41] although the essential details do not intrinsically require such. Chiral induction could also stem from other primordial biosynthetic reactions, and so the model we propose is not exclusive of alternative pathways. However, the RNA-centered path (supplementary file) is the most obvious candidate, given the limitations reflected by what is chemically reasonable under early primitive conditions.

Given that the parity non-conserving weak interaction is the only inherently handed property in nature,[42,43] we show it to be sufficient to differentiate between two enantiomeric forms in an autocatalytic reaction performed over extended periods of time (supplementary file), but only in the presence of catalytic metal ions available under primordial conditions, such as divalent alkaline earth metals. In particular, we detail calculations of the magnitude of the effect, the impact of various geologically-relevant metal ions, and the influence on evolution and dominance of chirality in the molecules of life. We establish a firm and realistic model that defines the prerequisites for the influence of the weak nuclear force to be evident in biochemical reactions and evolution through cellular pathways. While earlier work focused on the very small magnitude of the effect, and perhaps insignificance for understanding chirality in Nature, in fact with autocatalysis over relatively "brief" geological timeframes, and direct involvement of naturally abundant metal cofactors, the possible influence is demonstrable and highly probable.

In the case of the molecular system described herein (namely, a divalent metal mediating an aldol reaction of relevance to primordial RNA synthesis, (supplementary file)), the atomic center that experiences chiral discrimination through the weak nuclear force, and the center that is the source of the weak nuclear interaction, are distinct and are not directly connected via covalent bonding. Consequently, we do not use a molecular orbital treatment to develop a molecular wavefunction with contributions from each of the two centers, as used previously to address the single-center problem. Rather we consider direct overlap of carbon-centered orbitals with the nucleus of a divalent alkaline earth ion ($Ca^{2+}$, $Sr^{2+}$, or $Ba^{2+}$), each of which has been demonstrated to promote formaldehyde condensation to carbohydrate precursors.[44,45] This brings considerable simplification to calculations based on the use of equation (1),

$$\Delta E^{PNC} = \frac{Ga}{\sqrt{2}} \frac{Q_{\omega}(P)(\Delta)}{E_a - E_b} \qquad (1)$$

where $\Delta E^{PNC}$ is the second-order perturbation theory expression for the PNC-induced energy splitting, $G$ is the Fermi constant, $\alpha$ is the fine structure constant, $Q_{\omega}$ is the weak charge, $(P)$ is the PNC matrix element, $(\Delta)$ is the spin-orbit coupling term that gives the first-order correction to the wavefunctions considered, and $E_a$ - $E_b$ represents the energy difference between ground and higher energy wavefunctions for the spin-orbit coupling term.

Prior theoretical analysis has shown that for a single-center problem, or where both principal atoms in a two center system are identical, there is a $Z^5$ dependence on $\Delta E^{PNC}$, where Z is the atomic number (supplementary file). This is the origin of the heavy atom effect where elements of higher atomic number show enhanced $\Delta E^{PNC}$. However, as previously noted, in a single-center, the expected magnitude of $\Delta E^{PNC}$ is vanishingly small. In the situation described in this paper, two atom centers (carbon, C and an alkaline earth metal, M) are considered, where the (P) term is based on $M^{2+}$ and shows a $(Z_M)^4$ dependence, while the $Q_w$ component exhibits a $(Z_M)$ dependence, for an overall $(Z_M)^5$ dependence for $\Delta E^{PNC}$.

The energy difference resulting from the parity violating interaction at an atomic nucleus that we will define as carbon $C_\alpha$ (Fig. S2, supplementary file) is given by equation (1),[37,46] (and see also background details in the supplementary file). The PNC matrix element can be written in the form of equation (2),[37,46]

$$(P) = \frac{i\sqrt{3}}{4\pi} R_{ns}(0) \frac{dR_{n'p}(r)}{dr} |r = 0 \tag{2}$$

where $R_{ns}$ and $R_{n'p}$ refer to the radial wavefunctions of the relevant atomic orbitals.

The spin-orbit matrix element from carbon np orbitals is diagonal in np orbitals and can be determined by comparison to observed np coupling $\xi_{np}$, as defined by equation (3),[37,46]

$$(\Delta) = < n''p_x | f(\mathrm{r}) l_y | 2p_z > = \frac{i}{2} \xi_{2p} \tag{3}$$

or in the case of a high energy n"p orbital by direct computation by use of equation (3) and mixing of p-orbitals on the carbon bearing the nascent chiral center. The latter approach was taken in this study since a 7p orbital proved optimal in extending electron density from the prochiral carbon center $C_\alpha$ to the alkaline earth ion.

**Results**

In the context of PNC, catalysis by divalent calcium addresses several problems that have impacted prior studies in efforts to evaluate and measure the effect in atomic or molecular systems. Without loss of generality, these equations are initially applied to the case of calcium, the most abundant alkaline earth element in the earth's crust (Fig. S3, supplementary file), and later extended to barium and strontium. In performing these calculations, two simplifying features distinguish this work from prior estimates. First, the involvement of an alkaline earth cation ($Ca^{2+}$, $Sr^{2+}$, $Ba^{2+}$) and a carbon center immediately resolves the single-center problem. Second, the ionic (electrostatic) bonding between the metal cation and the organic molecule that is being acted on permits the problem to be tackled directly using atomic orbitals on calcium and carbon, since there is essentially no covalency or electronic sharing between each.

In defining the matrix elements, we take the representative case of a $Ca^{2+}$ ion separated from the carbon center of interest ($C_\alpha$ in Fig. S2, supplementary file), by a distance (a conservative upper limit) of 5.7 *Å*. This represents the internuclear distance for a linear arrangement of atoms extending from $Ca^{2+}$ to $C_\alpha$ with regular C-C and C-O covalent bonds between intervening carbon and oxygen centers and the ionic radius of $Ca^{2+}$.[47] This distance corresponds to 9.509 au in the case of $Ca^{2+}$, 9.849 au for $Sr^{2+}$, and 10.170 au for $Ba^{2+}$, and as noted earlier represents conservative upper limits on the key internuclear distance.

Given the ionic character of the bonding between calcium ion and the organic molecule, we further consider a direct through-space interaction between the orbitals on carbon and the calcium ion and then performed two sets of calculation. First, we assumed the volume integral to encompass only the volume of the nucleus of the alkaline earth ion. This evaluation yielded results for $\Delta E^{PNC}$ on the order of $< 10^{-20}$ au and too small to contribute detectable chiral selectivity, even over a geological timeframe. In a second series of calculations we assumed the volume element to include the ion. We believe the latter calculation to be best for the goal of interest, because the influence of the weak nuclear force is communicated to the nascent chiral carbon via an electronic interaction.

In this calculation, in spherical polar coordinates, we take a volume element of the carbon atomic orbitals around the sphere of the divalent ion determined by the ionic radius ($Ca^{2+}$ 2.155 au, $Sr^{2+}$ 2.495 au, $Ba^{2+}$ 2.820 au) at an inter-center distance representing the sum of the bond distances (both covalent and ionic) extending from the prochiral carbon to the charged ion ($Ca^{2+}$ 9.508 au, $Sr^{2+}$ 9.849 au, $Ba^{2+}$ 10.170 au). Angular variations for the θ and φ coordinates were evaluated from the geometric constraints of the internuclear distance and the ionic radii of each divalent ion, and consequently the angular terms involving $\int d\theta$ and $\int \sin\varphi \, d\varphi$ were each approximated as 0.25 using these limits. Simplifying assumptions were designed to provide conservative lower limits in subsequent calculations.

Values for various parameters defined in equations (1 to 3) were determined as follows. For calcium, electron density on the orbital (1s) most closely aligned with the atomic nucleus, was mixed with the lowest lying empty p-orbital (4p), and so $R_{1s}(0) = 2.(Z_{Ca})^{3/2}$. Using $Z_{eff}$ = 20 for electrons in the 1s orbital yields $dR_{4p}(r)/dr|_{r=0} = (10.(Z_{Ca})^{5/2})/32\sqrt{15}$, while $\sin^2\theta_w$ = 0.25 in equation S4 (supplementary file) yields $Q_w^{Ca}$ = -20 for $^{20}Ca$. Given the distance between the prochiral $C_\alpha$ atom and the divalent metal ion, the carbon 7p orbital provided the most significant contribution to the spin-orbit coupling term (Δ). The ΔE term in the denominator of equation (1) is estimated as 0.414 au, corresponding to the ionization energy for a carbon atom,[48-50] taking the ionization energy as a conservative upper limit for the energy difference between the carbon 2p and higher energy np orbitals and again representing a conservative limit on the magnitude of the possible energy differences. Table 1 summarizes the values of $\Delta E^{PNC}$ estimated from equation (1 to 3) with varying parameter values for each divalent cation and $C_\alpha$ carbon as indicated, and Table 2 illustrates how those $\Delta E^{PNC}$ values translate into enantiomeric selectivities over various time frames. In these calculations the temperature was taken as 315 K, representing an approximate average of ocean temperature on early earth.[51-53]

The influence of reaction rate constants was also examined, and summarized in Table 3. For the purpose of this article a modest rate constant of 1 $min^{-1}$ is taken for representative calculations, which approximates the activity reported for calcium promoted condensation of glycolaldehyde and formaldehyde (Figure S2).[45] More generally, Table 3 illustrates the impact of increasing or decreasing the rate constants by an order of magnitude (to 0.1 $min^{-1}$ and 10 $min^{-1}$).

**Table 1.** Summary of $\Delta E^{PNC}$ values determined for a single $Ca^{2+}$, $Sr^{2+}$, and $Ba^{2+}$ center, accounting for variations in orbital energies and atomic numbers. For $Ca^{2+}$, $Sr^{2+}$ and $Ba^{2+}$, n' = 4 to 6 orbital contributions, respectively, were considered for the parity term (P) in equation (2), and n"=3 to 12 contributions considered for the carbon spin-orbit coupling term (Δ) in equation (3).

| cation | $\Delta E^{PNC}$ (au) (one $M^{2+}$ catalytic ion) | $\Delta E^{PNC}$ (au) (two $M^{2+}$ catalytic ions) |
|---|---|---|
| $^{20}Ca^{2+}$ | $0.3 \times 10^{-20}$ | $0.6 \times 10^{-20}$ |
| $^{38}Sr^{2+}$ | $0.6 \times 10^{-19}$ | $1.2 \times 10^{-19}$ |
| $^{56}Ba^{2+}$ | $0.3 \times 10^{-18}$ | $0.6 \times 10^{-18}$ |

**Table 2.** Dependence of enantiomeric selectivities on $\Delta E^{PNC}$.

| $\Delta E^{PNC}$ (au) | selectivity factor over 4.2 B years | selectivity factor over 0.5 B years | selectivity factor over 0.2 B years | selectivity factor over 50 MM years | selectivity factor over 1 MM years | selectivity factor over $10^4$ years |
|---|---|---|---|---|---|---|
| $10^{-20}$ | 1 | 1 | 1 | 1 | 1 | 1 |
| $10^{-19}$ | 1 | 1 | 1 | 1 | 1 | 1 |
| $1.2 \times 10^{-19}$ | $6 \times 10^{12}$ | 33 | 4 | 1.4 | 1 | 1 |
| $5 \times 10^{-19}$ | $4 \times 10^{25}$ | $1 \times 10^{3}$ | 16 | 2 | 1.01 | 1 |
| $10^{-18}$ | $7 \times 10^{63}$ | $4 \times 10^{7}$ | $1 \times 10^{3}$ | 5.8 | 1.04 | 1 |
| $10^{-17}$ | $> 10^{300}$ | $3 \times 10^{68}$ | $2 \times 10^{27}$ | $7 \times 10^{6}$ | 1.4 | 1 |
| $5 \times 10^{-17}$ | $> 10^{300}$ | $> 10^{300}$ | $3 \times 10^{137}$ | $2.3 \times 10^{34}$ | 4.9 | 1.02 |
| $10^{-16}$ | $> 10^{300}$ | $> 10^{300}$ | $2 \times 10^{274}$ | $3.8 \times 10^{68}$ | 24 | 1.03 |
| $10^{-15}$ | $> 10^{300}$ | $> 10^{300}$ | $> 10^{300}$ | $> 10^{300}$ | $5.3 \times 10^{13}$ | 1.4 |
| $10^{-14}$ | $> 10^{300}$ | $> 10^{300}$ | $> 10^{300}$ | $> 10^{300}$ | $1.9 \times 10^{137}$ | 24 |
| $10^{-13}$ | $> 10^{300}$ | $> 10^{300}$ | $> 10^{300}$ | $> 10^{300}$ | $> 10^{300}$ | $> 10^{300}$ |
| $10^{-12}$ | $> 10^{300}$ | $> 10^{300}$ | $> 10^{300}$ | $> 10^{300}$ | $> 10^{300}$ | $> 10^{300}$ |
| $10^{-11}$ | $> 10^{300}$ | $> 10^{300}$ | $> 10^{300}$ | $> 10^{300}$ | $> 10^{300}$ | $> 10^{300}$ |
| $10^{-10}$ | $> 10^{300}$ | $> 10^{300}$ | $> 10^{300}$ | $> 10^{300}$ | $> 10^{300}$ | $> 10^{300}$ |
| $10^{-9}$ | $> 10^{300}$ | $> 10^{300}$ | $> 10^{300}$ | $> 10^{300}$ | $> 10^{300}$ | $> 10^{300}$ |
| $10^{-8}$ | $> 10^{300}$ | $> 10^{300}$ | $> 10^{300}$ | $> 10^{300}$ | $> 10^{300}$ | $> 10^{300}$ |

**Table 3.** Dependence of enantiomeric selectivities on rate constant for illustrative $\Delta E^{PNC}$.

| $\Delta E^{PNC}$ (au) | rate constant ($min^{-1}$) | selectivity factor over 50 MM years | selectivity factor over 1 MM years | selectivity factor over $10^4$ years |
|---|---|---|---|---|
| $10^{-17}$ | 10 | $3 \times 10^{68}$ | 24 | 1.03 |
| $10^{-17}$ | 1 | $7 \times 10^{6}$ | 1.4 | 1.003 |
| $10^{-17}$ | 0.1 | 4.8 | 1.03 | 1.0003 |
| $10^{-18}$ | 10 | $4 \times 10^{7}$ | 1.42 | 1.004 |
| $10^{-18}$ | 1 | 5.8 | 1.04 | 1.0004 |
| $10^{-18}$ | 0.1 | 1.2 | 1.004 | 1.00004 |

**Discussion**

While PNC has been previously suggested to play a role in chiral discrimination that differentiates the vast preponderance of D over L configuration ribose (and deoxyribose) in natural nucleic acids, and L over D amino acid configurations, the problem has been studied in isolation, with no clear path to incorporating such an effect into an evolutionary model. This requires chemical mechanisms that would allow discrimination to become manifest by natural biosynthetic pathways, and in a manner that would theoretically be large enough to produce an effect, even over geological timescales. In this paper we present such a model, based on the autocatalytic self-replication of glyceraldehyde as a precursor for simple sugars, and support its theoretical basis from calculations of the magnitude of the PNC effect from $Ca^{2+}$-promoted aldol coupling reactions (supplementary file). Most likely this is not the only pathway by which chiral discrimination might arise, but our work does provide a first step in exploring the scope of such an influence, and understanding the broader biochemical implications of the parity violating influence of the weak nuclear force in nature.

Prior solution kinetic studies of $Ca^{2+}$-promoted coupling of formaldehyde and glycoladehyde yielded rate constants of up to ~ 1 $min^{-1}$.[45] By use of equation (1) and assuming a calculated $\Delta\Delta E_A$ according to the alkaline earth ions summarized in Table 1, as determined from the influence of the PNC energy difference, the effective enantiomeric excess, as defined by the ratio of rate constants replicated over a given time period (equation (1)) can be readily calculated. Table 2 summarizes the enantiomeric ratio achieved with a range of $\Delta\Delta E_A$ and timeframes, with a fixed reaction rate constant, and with a fixed $\Delta\Delta E_A$ but variable rate constant and time frames. Assuming a rate constant of 1 $min^{-1}$ (or within the range of 0.1 – 10 $min^{-1}$ as reflected in Table 3) chiral discrimination is meaningfully evident well within evolutionary time frames, and even over only 50 to 100 million years when the earliest life forms are believed to be evolving.[54] Taking a selectivity factor of 1.4 as an arbitrary and illustrative measure over a time frame of $10^4$ years, a $\Delta\Delta E$ threshold of ~ $10^{-15}$ au appears to be a reasonable estimate of the energy difference required, $10^{-17}$ au for a 1 million year period, and 1.2 x $10^{-19}$ au if one considers a 50 million year period.

As noted earlier, temperature is a factor that also influences kinetic activation barriers (as represented by equation S1). Ocean temperatures of early earth have been estimated from O/Si isotope ratios and nucleic acid melting temperatures from ancient archaebacteria,[51-53] and found to vary from ~ 343 K initially and cooling to 290 K. The high and low limits, were considered, with slightly lower enhancement as temperature increases, although these are offset to some extent by more rapid catalytic turnover. However, in the time range of interest the difference in temperature does not change the results significantly, and certainly not the key conclusions.

While divalent calcium is the most likely cation to promote catalytic activity, based on natural abundance (Fig. S3, supplementary file) and prevalence in primordial clays and minerals, nevertheless heavier elements such as barium and strontium that show an enhanced Z- effect should also be considered. Calculations for each metal ion show a significant enhancement in the magnitude of the $\Delta\Delta E$ influence from the weak nuclear force (Table 1), accounting for differences in Z and $Q_W$. The timeframe required for a significant effect is also reduced if the heavier elements are implicated (Table 2). Both barium and strontium occur in relatively high concentration in the earth's crust (Figure S3) and are also prevalent in many calcium minerals (such as barytocalcite, $BaCa(CO_3)_2$, and olekminskite, $Sr(Sr,Ca,Ba)(CO_3)_2$) in addition to their own mineral forms.

These enhancements notwithstanding, the aim of this work was to re-establish the relevance of parity violation as a likely mechanism for the origin of molecular chirality in the molecules of life. Such a view has been criticized based on the magnitude of the effect and the lack of viable mechanisms for chiral discrimination to be manifest.[29,55] In this paper we have addressed both of these concerns and demonstrate that a combination of metal-promoted catalysis and viable chemical pathways can readily account for chiral selectivity in the context of the RNA world model over evolutionary timeframes.

While Earth is estimated to be around 4.5 billion years old it is unreasonable to take that entire timeframe over which to establish a chiral preference. Estimates of when life first appeared on earth vary, but evidence suggests the appearance of simple bacteria most likely occurred ~ 3.5 to 4.0 billion years ago, providing a time window of 0.5 to 1.0 billion years for chirality to emerge in the fundamental building blocks of life. Given the aforementioned kinetic criteria and timeframes, an energy difference exceeding $10^{-19}$ au is desirable. While this places $Ca^{2+}$ at the edge of the limit for a chirality-selecting catalytic cofactor, both $Sr^{2+}$ and $Ba^{2+}$ lie very much within the realm of feasibility. Given that conservative estimates have been used in all calculations, and the sensitivity of chiral selectivity to rate constants and especially calculations of $\Delta E^{PNC}$, it would be unwise to completely discount a possible role for $Ca^{2+}$ in chiral discrimination.

Other factors that contribute to the energy difference include the identity of the catalytic metal cofactor, the number of metal cofactors involved, and variations in rate constants for the coupling reactions in Figs S1 and S2 (supplementary file), but the goal of this work is directed toward establishing the viability of $\Delta E^{PNC}$-related mechanisms to achieving chiral selectivity, and biologically-relevant autocatalytic chemistry as a framework on which to build that understanding. We also do not wish to discount other viable alternatives. It is entirely possible that chirality in the sugar molecules defined herein, or that chiral preference in other completely distinct classes of biomolecules such as amino acids could have evolved by one or more of the distinct mechanisms outlined earlier. One of these pathways could dominate, or collectively two or more independent chemical paths mediated by distinct chiral centers could work together in a synergistic fashion, either contemporaneously, or in distinct time frames. Alternatively, seeding mechanisms promoted by chiral materials from extraterrestrial meteorites and comets[8,9] could initiate chiral selectivity, depending on the specific chirality of the “seed molecule”. A possible test of the latter could emerge from evidence of preferred handedness of molecules on exoplanets. If parity violation is not the primary factor then evaluation of preferred chiral forms on a large number of exoplanets should reveal a statistical 50:50 split. But if parity violation is the key factor then an earth-like preference should be uniformly observed.

Resolution of these broader questions must await appropriate technological advances in assessing chirality in distant planetary systems. However, the model described in this work does provide an autocatalytic mechanism that lies in the early stages of an RNA-centered world view. It uniquely offers mechanistic features and possibilities that are not readily envisaged in other proposed schemes, demonstrates how the requirement for metal ion catalysis moves the PNC effect to a position of relevance in addressing this problem, and is fully consistent with available data. We hope this analysis spurs additional work in this area of fundamental research.

**Acknowledgements**

We extend thanks to Drs. Sherwin Singer and Dehua Pei (The Ohio State University) for critical reading of the manuscript and helpful suggestions.

## Supplementary File

**Origin of Chirality in the Molecules of Life**

J. A. Cowan and R. J. Furnstahl

Contribution from the Department of Chemistry and Biochemistry, The Ohio State University, 100 West 18th Avenue, Columbus, Ohio 43210, and the Department of Physics, The Ohio State University, 191 West Woodruff Avenue, Columbus, Ohio 43210.

The following information on the biochemical, kinetic, and theoretical aspects of this work are presented as a guide of readers requiring additional background information on some of the topics covered.

**Autocatalytic Chemistry Relevant to Formation of Primordial RNA**

The RNA world model posits the role of self-replicating RNA as a basis for the later evolution of DNA and proteins.[1,2] A prerequisite is the formation of the basic components of RNA, which includes the ribose ring. The ribose sugar contains the only molecular chiral center in ribonucleic acid and is therefore an attractive progenitor of chirality from RNA catalytic formation of other molecular components that stem from such a model.

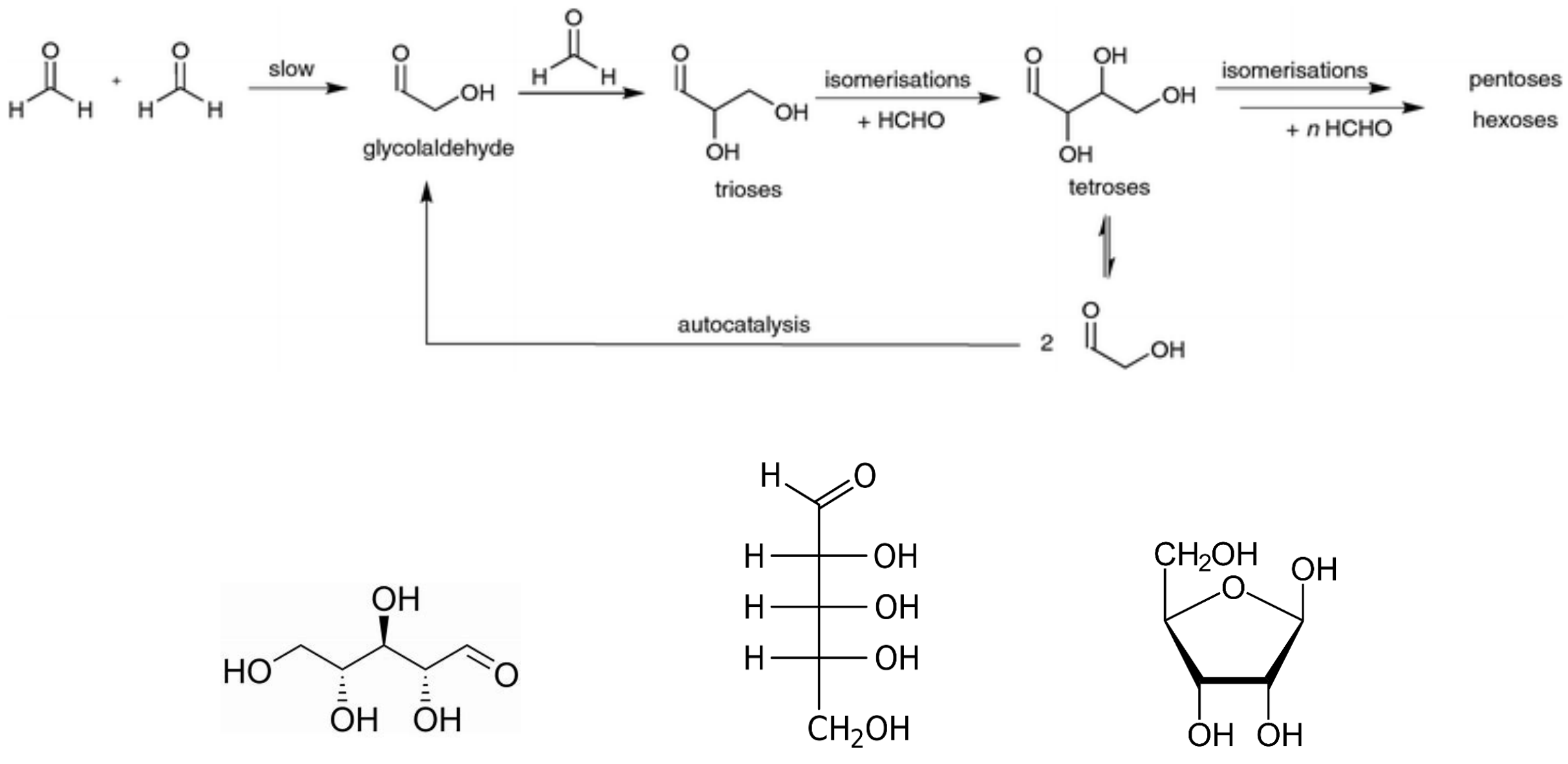

**Fig. S1.** (top) Autocatalytic pathway to ribose via glyceraldehyde. (bottom) Both ring open and closed representations of ribose, and a Fisher projection (center) are shown.

The formose reaction (Fig. S1), discovered in 1861 by Butlerow,[3] is an autocatalytic reaction that produces sugars from the simple organic molecule formaldehyde ($CH_2O$), involving a series of aldol and reverse aldol reactions and isomerizations. Formation of glyceraldehyde by coupling of formaldehyde and glycolaldehyde yields a stereocenter ($C_\alpha$, Fig. S2). Breslow first outlined an autocatalytic pathway that provides key building blocks for formation of sugars prevalent in nucleotides.[4-6] Both condensation and isomerization reactions are catalyzed by divalent calcium, an element that would have been particularly prevalent in primitive clays and minerals (Fig. S3).[7] Experimental work that reproduced an environment similar to that expected in primordial times, and including mineral forms such as colemanite ($Ca_2B_6O_{11}5H_2O$) and kernite ($Na_2B_4O_7$),[8] demonstrated formation of ribose, the pentasaccharide component of RNA. Interestingly, spectroscopy has revealed the presence of formaldehyde and glycolaldehyde in space, however, it is unclear if the chemistry and autocatalytic processes described in this work are relevant to selective formation of stereocenters in astrobiology.

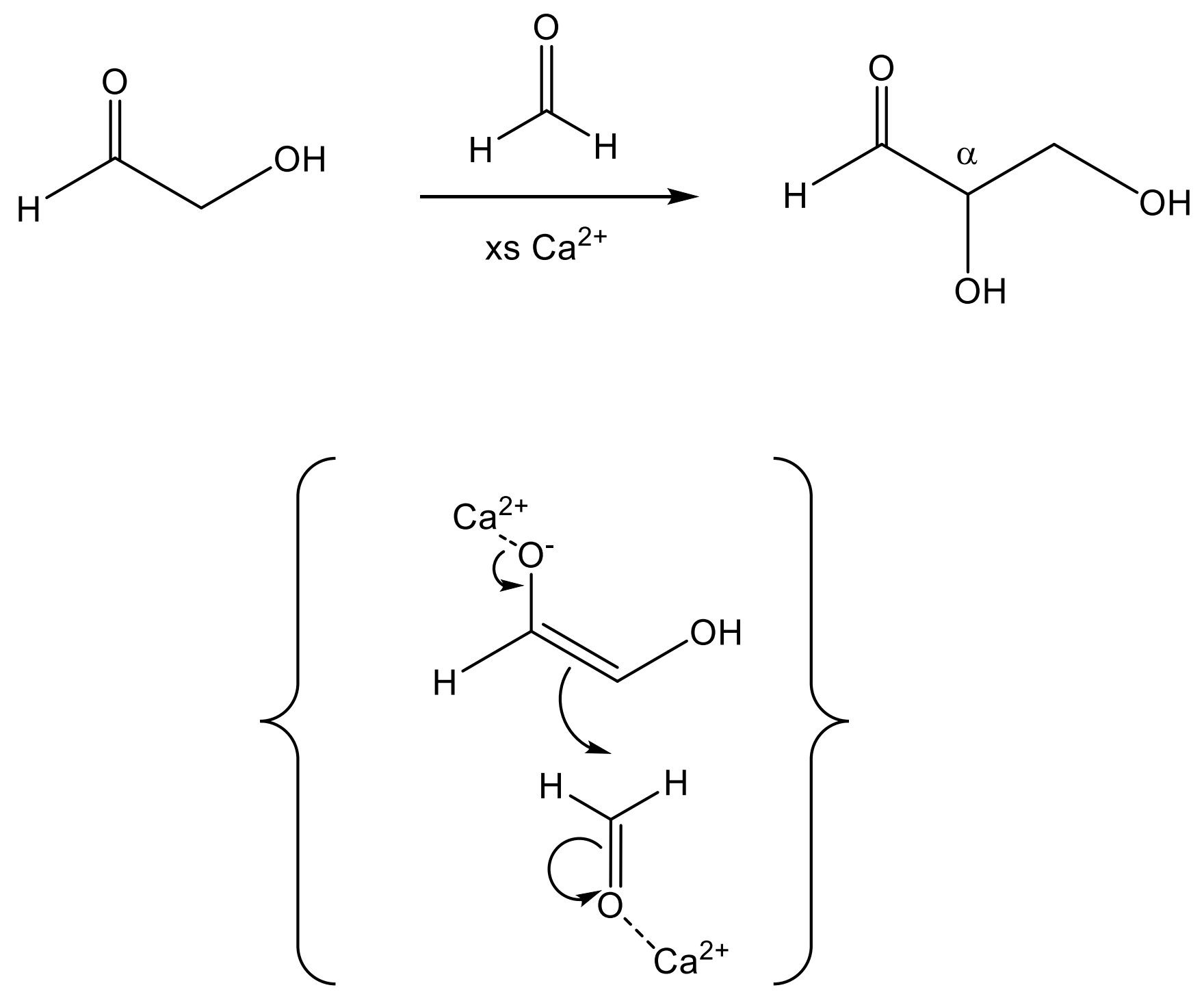

**Fig. S2.** Glyceraldehyde formed by coupling of glycolaldehyde and formaldehyde via an aldol reaction

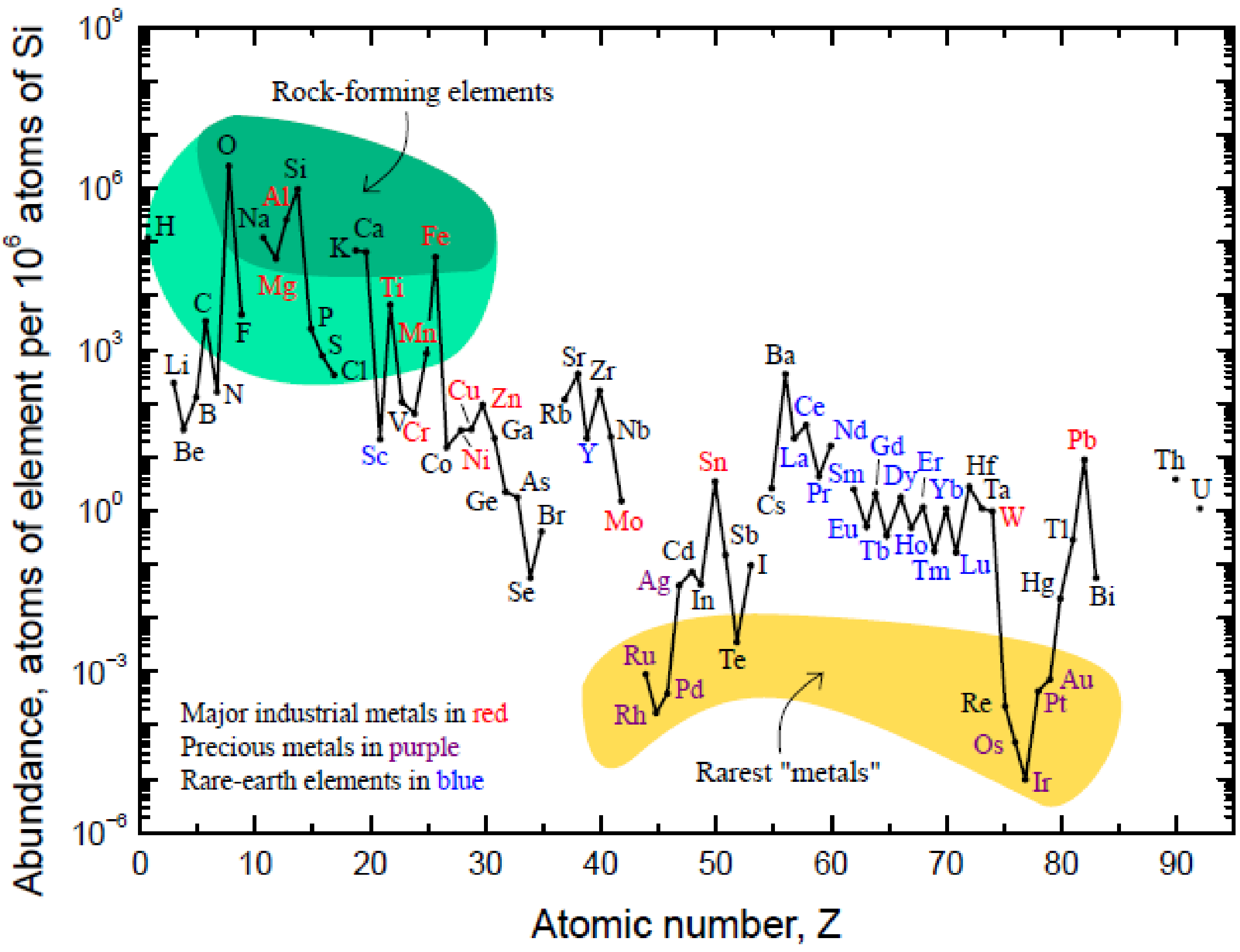

**Fig. S3.** Relative abundance of elements in the earth's crust. Taken from reference 7.

**Chemical Kinetics and Autocatalysis**

Discrimination in the formation of two enantiomeric forms can arise from distinct activation energies reflecting diastereomeric discrimination in the transition state (Fig. S4). Without loss of generality, it is assumed that $\Delta E^*_1 > \Delta E^*_2$. For the subject of this paper, the difference in activation energies ($\Delta\Delta E^*$) will be extremely small, however, in the context of an autocatalytic reaction that is repeated multiple times, and in particular over evolutionary timescales, the enhancement of one enantiomer over another will be significant and is reflected by equation (S1), where n is the number of catalytic turnovers within the time period under consideration, based on the experimental rate constants of ~ 1 $min^{-1}$ for $Ca^{2+}$-promoted coupling of formaldehyde and glycoladehyde.[9]

$$k^{\text{enhance}} = (exp(\Delta\Delta \mathrm{E} */\mathrm{RT}))^{n} \qquad \text{(S1)}$$

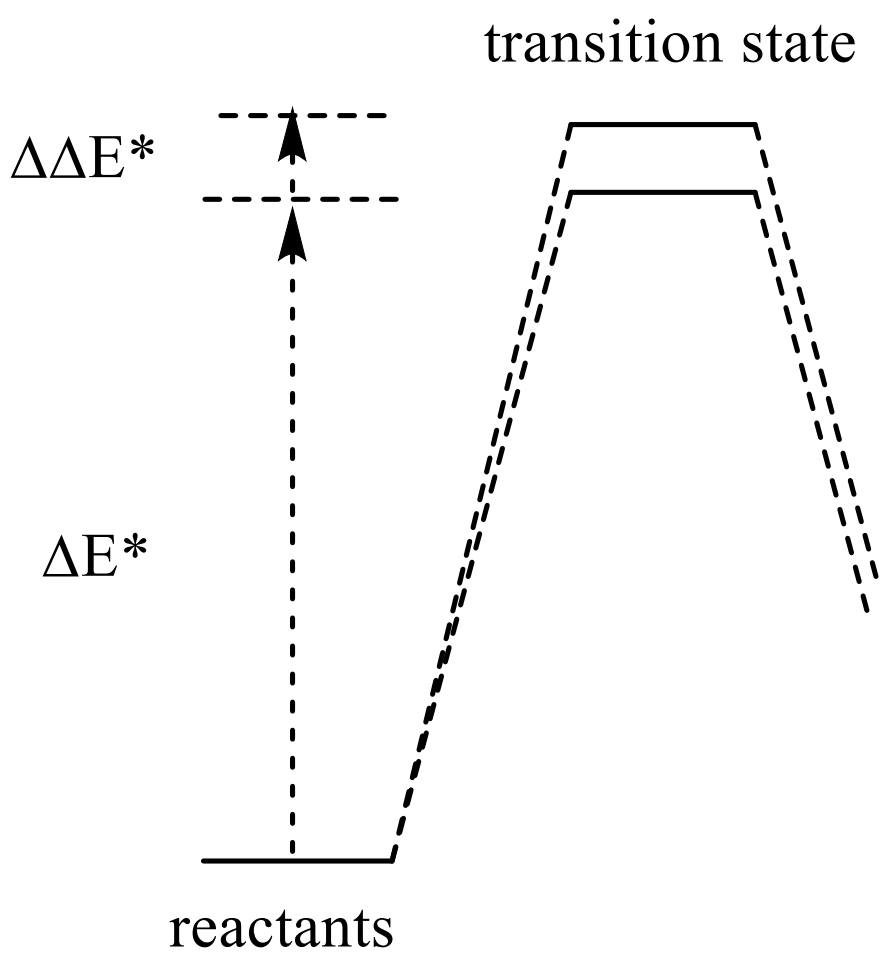

**Fig. S4.** Activation barriers for reactivity, where $\Delta\Delta E^* = \Delta E^{PNC}$.

**Weak Nuclear Force and Parity Violation**

Since the proposal by Yang and Lee in the 1950's that the weak nuclear interaction does not obey the law of parity conservation,[10] and experimental validation by Wu,[11] it is now well established that this represents the only such force in nature. Advancement of the theory to understand how the weak nuclear force might influence molecular properties (both chemical and physical) builds on extensive theoretical developments that initially focused on atomic properties[12,13] and later evolved to molecular systems.[14-19] Of particular note is the pioneering work by Bouchiat and Bouchiat that highlighted the potential to observe PNC effects in atomic spectroscopy,[12,13,20] although for reasons to be discussed, the expected magnitudes were small to non-measureable.

While combinations of electron-electron, nucleon-nucleon and electro-nucleon interactions should be considered, the latter is dominant, and the only term generally incorporated when considering potential "observable" effects (equation S2).

$$V_{eA}^{PNC} = \frac{G\alpha}{4\sqrt{2}} Q_{\omega}^{\alpha} \{\sigma_i \cdot p_i, \rho_N(r)\}_+ \quad \text{(S2)}$$

Dirac delta functions are centered on nucleus α for each electron i (equation S3),

$$V^{PNC} = \frac{G\alpha}{4\sqrt{2}} \sum_{\alpha,i} Q_{\omega}^{\alpha} \{\sigma_i . p_i , \delta^3(r_{i\alpha})\}_+ \quad = \quad \sum_i v_i^{PNC} \tag{S3}$$

where $\rho_N(r)$ is a delta function on the scale of an electron wavefunction.

For many electron atoms, the interactions are summed over the electrons and nuclei and where the weak charge $Q_\omega$ is defined by equation (S4).

$$Q_w^{Ca} = (1\text{-}4\sin^2\theta_\omega) - N \tag{S4}$$

Among several active proponents of the important influence of the weak nuclear force and parity nonconservation (PNC) on chiral discrimination in molecular systems, the work of Hegstrom is of particular importance,[18,21] establishing a benchmark for later studies. In these studies both a PNC interaction potential and the molecular wavefunctions on which it acts were viewed as important contributors to defining the overall magnitude of the effect. Hegstrom described a thorough analysis of the problems and requirements associated with a more rigorous use of molecular wavefunctions and orbitals in application to specific molecules, and in particular delineated the requirements for non-vanishing contributions to the expectation value to remain. Since the wavefunctions of closed shell systems are singlets and real they give rise to no effective value of the weak PNC potential. The latter is imaginary and dependent on spin-orbit coupling, and so in a non-relativistic treatment there is a need for an additional spin-orbit coupling term in the overall energy equation.

Hegstrom highlighted the "single-center" problem, an important restriction that undermines manifestation of the PNC effect in both atomic and molecular systems where the effect was focused on a single atom. It has been shown by the single-center theorem that the influence on the electronic structure of an atom by the same atomic nucleus is essentially zero. In brief, the δ function in equation (S3) has non-zero matrix elements only between s and p states, while those from the spin-orbit coupling matrix elements are only non-zero for orbitals other than s ($l$ = 0). In this case, as demonstrated by Hegstrom et al, $\Delta E^{PNC} = 0$ if both matrix elements refer to orbitals or states on the same nucleus. For this reason, the magnitude of $\Delta E^{PNC}$ is greatly increased if the two matrix elements are centered on different atoms.

The effect is greatly enhanced if the weak nuclear force stems from an adjacent and distinct atomic nucleus, and acts on the electronic structure of the "prochiral" nucleus of interest, which in our case is the carbon designated $C_\alpha$ (Fig. S2).

Another important caveat is the dependence on atomic number (up to $Z^5$ for a given nucleus) that can further limit the observability of the PNC effect to heavier nuclei. For this reason, very early work soon focused on heavy elements such as uranium, to accentuate an effect that was already made challenging by the single-center theorem.